\documentclass[twocolumn,showpacs,preprintnumbers,amssymb]{revtex4}
\usepackage{graphicx}% Include figure files
\usepackage{dcolumn}% Align table columns on decimal point
\usepackage{bm}% bold math

\begin{document}
\input epsf

\title{Non-adiabatic dissociation of molecules and BEC loss due to shock-waves}
\author{Nir Gov}
\address{Department of Materials and Interfaces,
The Weizmann Institute of Science,\\
P.O.B. 26, Rehovot, Israel 76100}

\begin{abstract}

Recent experiments have shown the likely appearance of coherent
BEC atom-molecule oscillations in the vicinity of a Feshbach
resonance. In addition, a new loss mechanism was observed, whereby
the loss of atoms from the BEC is inversely dependent on the rate
of change of the applied magnetic field. We present here a
phenomenological model which gives a good description of the
scaling properties of this new decay process, by attributing it to
non-adiabatic dissociation of molecules by a propagating
shock-wave. The model has only two free parameters, which specify
the size of the "shocked-region", and can be readily tested by
future experiments.

\end{abstract}


\maketitle

In a recent series of papers the coherent creation of molecules
from an atomic BEC close to the Feshbach resonance, was
demonstrated \cite{nature,prl02}. Both sets of experiments have
the following principle characteristics. In these experiments a
stable $^{85}$Rb BEC (at initial field $B_{i}$) was subjected to a
well controlled linear magnetic field pulse which brought it over
a certain rise time $t_{r}$ to a final field $B_{f}$, close to the
critical field $B_0\simeq154.9$G of the Feshbach resonance. After
a certain holding time $t_{h}$ the field was ramped back to the
initial value and the BEC was imaged. There was the distinct
periodic appearance of a hot ($\sim$150nK) burst of atoms, and an
unseen component, all of which remained coherent with the atomic
BEC component. The unseen component is presumably the molecular
component, since the Rabi frequency for the periodic change agrees
with the energy difference between the free atoms and the bound
molecular level $E_{am}(B_{f})$ \cite{nature,holland}.

In addition to the periodic decrease and revival of the BEC, the
results indicate an new mechanism of loss, which is seen to be
much faster than the atomic two- and three-body inelastic loss
rates \cite{2body}. All attempts to explain this loss in terms of
adiabatic dynamics, treating the BEC as isotropic and uniform,
have not been successful \cite{prl02,fin}. It is therefore useful
to try a different approach. The decrease in the atomic BEC could
be due to molecules being formed or dissociated non-adiabatically,
so that they do not take part anymore in the coherent oscillations
\cite{nature}. One possible non-adiabatic mechanism is the
creation and propagation of shock waves in the atomic-molecular
cloud. Shock waves in a classical gas are supersonic
discontinuities in the temperature, entropy, pressure and density
\cite{landuahydro}, with all these quantities being increased by
the passing shock-wave. Similar dynamic discontinuities in the
trapped cloud can result in non-adiabatic dissociation of
molecules and therefore loss of coherence of the resulting atoms
with respect to the atomic BEC. It is this process which we shall
describe using a phenomenological model here. We will not attempt
a full simulation of the dynamics of the atomic and molecular
clouds, but will describe the scaling of BEC loss due such a
process.

We begin by giving a qualitative description of mechanisms which
can cause the appearance of shock-waves in the trapped cloud, due
to the rapid change in the applied magnetic field. When the
magnetic field is changed from its initial value $B_{i}\sim 166$G
to its final value $B_{f}$ close to the Feshbach resonance, the
elastic atomic scattering length $a$ changes by a factor of up to
\cite{prl02} $\sim4000$, given by \cite{nature}
\begin{equation}
a=a_{bg} \left(1-\frac{\Delta}{B-B_{0}} \right)
 \label{ab}
\end{equation}
with $a_{bg}=-450a_{0}$, $a_{0}=0.053$nm and $\Delta=11$G.

The equilibrium value of the density, using the usual Thomas-Fermi
limit \cite{bec}, is: $n_{eq}\propto a^{-3/5}$. The BEC therefore
finds itself with the initial density $n$, which is much larger
than the new equilibrium density $n_{eq}$ that corresponds to the
new scattering length $a(B_{f})$. This means that the BEC cloud
wants to greatly expand in response to the increase in $a$. The
new equilibrium chemical potential $\mu_{eq}=4\pi h^2 n_{eq}
a/m\propto a^{2/5}$ is also very different from the chemical
potential at which the cloud finds itself prior to any expansion,
i.e. $\mu=4\pi h^2 n a/m$.

The expansion of the cloud to the new equilibrium size happens
over a relatively long time scale, given by the largest trap
frequency, which in the experiments is $\tau_{trap}\simeq30$ms.
But even before the cloud has time to readjust, the density begins
to change through the propagation of density (sounds) waves,
similar to the case of an expanding classical gas. Rarefaction
waves at the BEC surface converge on the axis of symmetry and
reflect as a shock-wave (Fig.1). Referring to a detailed
simulation of a similar dramatic change in the scattering length
\cite{saito}, we note that the imploding dynamics of the BEC
create localized spikes of density (estimated to be up to $10n$),
of lateral size \cite{saito} $W\sim 0.1\mu$m. The width of the
"shocked region" $W$ is an essential parameter in our model
(Fig.1).

Another source of perturbation which can cause the shock-waves to
appear, is the coherent conversion of atoms into molecules. This
process will not be uniform in amplitude, being denser in the
middle of the cloud, or in its frequency, due to the different
local magnetic field across the cloud. The BEC will therefore have
a nonuniform density of molecules, oscillating (on average) at the
Rabi frequency \cite{nature,holland} $10-40$kHz, depending on the
holding field $B_{f}$. This time scale of $100-20\mu$sec is of
order of the time-scale of the observed loss rate.

Our premise is therefore the following: when the magnetic field is
suddenly changed, acoustic waves propagate in the cloud in an
attempt to restore the uniformity of the chemical potential, which
is nonuniform due to the density profile of the cloud.
Additionally, the BEC will be perturbed by an oscillating and
nonuniform density of molecules. The consequent fluctuations
(waves) in density will result in reverberating shock-waves. Note
that by the term shock-waves, we mean in general a propagating,
non-adiabatic excitation.

We estimate the amplitude of the shock-waves to be proportional to
the gradients of chemical potential which form across the BEC due
to the finite time it takes an acoustic wave to traverse the
relevant region of the cloud. The discontinuity in the chemical
potential at the shock front $\Delta \mu$, is therefore estimated
to be the accumulated change in $\mu$ over the time $\tau=W/v_{s}$
it takes a sound wave to cross the "shocked-region" of size $W$,
with the velocity of sound given by the standard expression
\cite{bec}: $v_{s}=\sqrt{\mu/m}$ (where we use the average density
$\langle n \rangle$ and the velocity of sound at the holding field
$B_{f}$). Note that weak shock-waves propagate at the speed of
sound \cite{landuahydro}. The chemical potential change,
accumulated over the time $\tau$ is
\begin{equation}
\Delta \mu(\tau)=\mu(\tau)-\mu_{0}\simeq 4\pi h^2 n a(\tau)/m
\label{mutr}
\end{equation}
(where $\mu_{0}=4\pi h^2 n a_{0}/m$, $a_{0}=a_{bg}
\left(1-\Delta/(B_{i}-B_{0})\right)\simeq-2$\AA), where
\begin{eqnarray}
a(\tau)&=&a_{bg} \left(1-\frac{\Delta}{B(\tau)-B_{0}} \right) \nonumber \\
B(\tau)&=&\left\{
\begin{array}{lc}
B_{f}+(B_{i}-B_{f})\frac{t_{r}-\tau}{t_{r}} & \; (t_{r} \geq \tau)
\\
B_{f} & \; (t_{r}<\tau)
\end{array}
\right.
 \label{abtr}
\end{eqnarray}
describing the linear change in the magnetic field with time
\cite{prl02}. The dependence on the rise time $t_{r}$ is therefore
clear: the faster the rise, the larger are the chemical potential
gradients created in the "acoustic" time $\tau$, and the resulting
amplitude of the shock-waves.

Note that our treatment is inherently dependent on the
inhomogeneity of the trapped cloud (Fig.1); in an infinitely
uniform cloud there will be no shock waves. The process of
shock-wave creation, even though we do not describe it here
explicitly, depends on inhomogeneities which propagate in the
cloud. We further immediately conclude that the decay time-scale
$\tau$ is largely independent on the BEC density $n$, as was found
\cite{prl02}. This behavior arises if the width of the
"shocked-region" $W$ and its length, are proportional to the
equilibrium radius of the initial BEC, which is $\propto n^{1/2}$,
while the sound velocity is also $v_{s}\propto n^{1/2}$
\cite{bec}.

The sound velocity is given by the magnetic field during the hold
time, i.e. closest to the Feshbach resonance. For $B_{f}=156.7$G
we have $v_{s}\simeq8$mm/sec, for initial BEC density $\langle n
\rangle=1.9\times 10^{13}$cm$^{-3}$ \cite{prl02}. To compare our
calculation with the measured data \cite{prl02}, we used the
measured decay time \cite{prl02}: $\tau=W/v_{s}=13.2\mu$sec, which
corresponds to $W\simeq 0.1\mu$m.

We now propose that these shock waves excite the molecules
non-adiabatically and create pairs of atoms, which are therefore
lost from the BEC (a similar mechanism of molecular
excitation/decay was mentioned in \cite{holland}). For the
dissociation probability per molecule, due to the passing
shock-wave, we assume a Boltzman-like factor:
$p(\tau)=\exp(-E_{am}/\Delta \mu(\tau))$. In a classical gas a
passing shock front heats the medium, and the energy jump at the
front, $\Delta \mu$, plays the role of effective temperature
\cite{landuahydro}. Assuming that the shock-waves propagate over
the width $W$ at constant speed $v_{s}$, and cover a constant
fraction $f$ of the cloud (i.e. the "shocked-region"), the rate of
loss of BEC is given by
\begin{eqnarray}
\frac{dN_{f}}{dt}&=&-\frac{N_{f}p(\tau)}{\tau} \nonumber \\
&\Rightarrow & N(t_{r})/N_{0}=f e^{-t_{tot} p(\tau)/\tau}+(1-f)
\label{fracremain}
\end{eqnarray}
where we took the total time for the shock-wave depletion process
to be empirically $t_{tot}=t_{h}+t_{r}/4$ \cite{prl02}, $N_{f}$ is
the number of atoms in the "shocked region", which initially is
equal to $fN_{0}$. The parameter $f$ is introduced in order to
take into account the observation that the loss mechanism we
describe here affects only a finite fraction $f\simeq90-60\%$ of
the cloud (Fig.2 of \cite{prl02}). A possible reason for this
behavior could be that the elongated shape of the cloud makes the
conditions for effective shock-wave creation and subsequent
molecular dissociation appear only along its central part (Fig.1).
This effect could explain why the proportion of the cloud which is
affected by this loss mechanism is smaller in the case of the
cloud with the smaller density \cite{prl02}. This BEC will be
shorter and therefore have a proportionately shorter central
"shocked" region. The two free parameters of our model are
therefore the width $W$ (which determines $\tau$) and the relative
number of atoms $f$ in the "shocked-region", out of the total
number of atoms in the cloud.

To take into account the background processes of two- and
three-body inelastic loss rates \cite{2body}, we next multiply
$N(t_{r})/N_{0}$ (\ref{fracremain}) by:
$\exp{[-t_{h}/\tau_{2-3}]}$, with a typical decay time of
$\tau_{2-3}\simeq0.12\times 10^{-2}$sec (in the vicinity of the
Feshbach resonance \cite{2body}).

In Fig.2 we compare our calculated fraction of remaining BEC
$N(t_{r})/N_0$ (Eq.(\ref{fracremain})) as a function of $t_{r}$,
for different hold times $t_{h}$, with the experimental data
\cite{prl02}. In the limit of vanishing rise time $t_{r}$, the
fraction of the remaining BEC shows an exponential decay with the
hold time $t_{h}$ (Fig.2 of \cite{prl02}). For increasing rise
time we first see the increasing loss, as long as $t_{r}<\tau$,
simply due to the increase in the overall time spent close to the
Feshbach resonance. As soon as $t_{r}>\tau$, the loss is
decreased, due to weakening shock-wave amplitude
(\ref{mutr},\ref{abtr}). The overall agreement is very good. The
discrepancy at the longest holding time ($t_{h}=100\mu$sec) arises
from the fact that the reverberating shock-waves weaken with time,
making longer holding times relatively less effective. This effect
is not taken into account in Eq.(\ref{fracremain}), where we
assumed that the dissociation probability $p(\tau)$ is constant
with time. We are therefore overestimating the loss of BEC for the
longer holding times. A much better agreement is achieved for the
case of $t_{h}=100\mu$sec, if an effectively reduced value of
$t_{h}\sim50\mu$sec is used in the calculation (\ref{fracremain}),
describing the decay of the shock-waves (dashed-line in Fig.2).
The background decay due to two- and three-body processes
\cite{2body} is shown by the dotted line.

In Fig.3 we compare our calculation (\ref{fracremain}), for a
constant holding time $t_{h}=1\mu$sec (Fig.4 of \cite{prl02}), and
varying final field $B_{f}$. The velocity of sound, and therefore
the decay time $\tau$, both depend on the magnetic field $B_{f}$,
with $\tau$ decreasing as we approach the resonance field $B_0$.
Additionally, the dependence of the dissociation energy $E_{am}$
on the magnetic field was previously measured \cite{nature}. Again
we find a good agreement, except for the longer rise times and
closest approach to the resonance. At this field, our empirical
assumption of $t_{r}/4$ being added to the overall decay period is
questionable, with a larger proportion probably active. On the
other hand, this means that the shock-waves have decayed for a
longer time too, resulting in our overestimation of the BEC loss,
as discussed above in relation to Fig.2.

To conclude, we have presented a phenomenological model which
attributes the loss of atomic BEC to non-adiabatic dissociation of
molecules by imploding shock-waves, created as a result of the
rapid change in the magnetic field. This simple model appears to
capture the main physical mechanism at work, as it describes the
correct dependence of the observed BEC loss on the various
physical parameters. Rigorous numerical simulations are needed in
order to substantiate this proposal. It could be interesting to
test this model further by repeating the experiments with BEC
clouds of different geometries (e.g. prolate vs. oblate), and by
producing shock-waves in controlled regions of the cloud by
applying time-dependent and spatially non-uniform electro-magnetic
fields.

\begin{acknowledgments}
I thank Ehud Altman, Roee Ozeri and Nir Davidson for useful
discussions. This work was supported by the Louis L. and Anita M.
Perlman Postdoctoral Fellowship.
\end{acknowledgments}

\begin{figure}
\centerline{\ \epsfysize 7cm \epsfbox{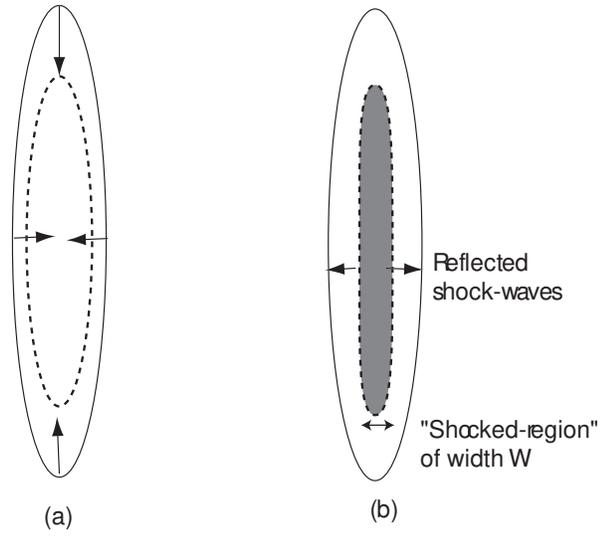}} \vskip 3mm
\caption{A schematic illustration of our proposed scenario,
whereby (a) imploding acoustic (rarefaction) waves are (b)
converging on the axis of symmetry to produce a central
"shocked-region" (dark region), where the non-adiabatic molecular
dissociation takes place.}
\end{figure}

\begin{figure}
\centerline{\ \epsfysize 7cm \epsfbox{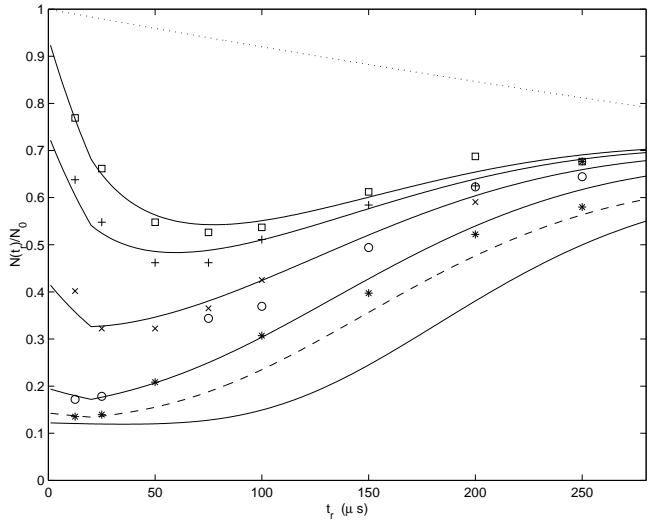}} \vskip 3mm
\caption{The calculated (solid lines, Eq.(\ref{fracremain})) and
measured \cite{prl02} (symbols) remaining fraction of atomic BEC
as a function of the rise time $t_{r}$, for different holding
times $t_{h}=1,5,15,35,100\mu$sec (from top to bottom). The dotted
line gives the background decay due to two- and three-body
processes \cite{2body}. The dashed line describes the calculation
for a holding time of $t_{h}=50\mu$sec.}
\end{figure}

\begin{figure}
\centerline{\ \epsfysize 7cm \epsfbox{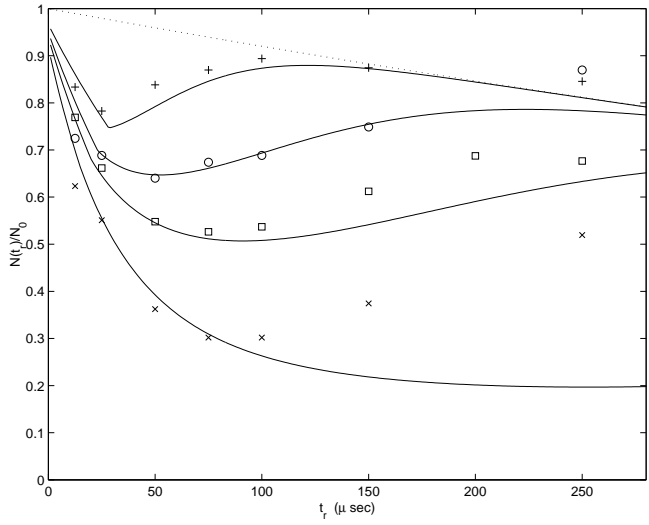}} \vskip 3mm
\caption{The calculated (solid lines, Eq.(\ref{fracremain})) and
measured \cite{prl02} (symbols) remaining fraction of atomic BEC
as a function of the rise time $t_{r}$, for different holding
magnetic field $B_{f}=158,157.2,156.7,156$G (from top to bottom).
The dotted line gives the background decay due to two- and
three-body processes \cite{2body}.}
\end{figure}

\end{document}